# Data sharing practices across knowledge domains: a dynamic examination of data availability statements in *PLOS ONE* publications

Chenyue Jiao (School of Information Sciences, University of Illinois at Urbana-Champaign, IL, USA)

Kai Li (School of Information Resource Management, Renmin University of China, Beijing, China)

Zhichao Fang (Centre for Science and Technology Studies, Leiden University, the Netherland)

Corresponding author: Kai Li, kai.li@ruc.edu.cn

**Abstract**

As the importance of research data gradually grows in sciences, data sharing has come to be encouraged and even mandated by journals and funders in recent years. Following this trend, the data availability statement has been increasingly embraced by academic communities as a means of sharing research data as part of research articles. This paper presents a quantitative study of which mechanisms and repositories are used to share research data in *PLOS ONE* articles. We offer a dynamic examination of this topic from the disciplinary and temporal perspectives based on all statements in English-language research articles published between 2014 and 2020 in the journal. We find a slow yet steady growth in the use of data repositories to share data over time, as opposed to sharing data in the paper and/or supplementary materials; this indicates improved compliance with the journal's data sharing policies. We also find that multidisciplinary data repositories have been increasingly used over time, whereas some disciplinary repositories show a decreasing trend. Our findings can help academic publishers and funders to improve their data sharing policies and serve as an important baseline dataset for future studies on data sharing activities.

# 1 Introduction

As data emerges as one of the most fundamental objects in scientific research, the flow of research data has become a central condition in contemporary research practices in nearly every knowledge domain. This condition, a tenet of the growing open science movement, requires data to be sharable to the broader community [1,2]. Major benefits of sharing data have been well documented in the literature, including improved scientific reproducibility and data reusability and fairer assignment of credit to researchers who contribute to science in different ways [3,4]. As part of this trend, major funders, including the National Institutes of Health and the National Science Foundation, have encouraged or even mandated data sharing since the 2000s [5,6]. This has greatly promoted data sharing among researchers [4,7].

However, the implementation of these policies is far from consistent across knowledge domains and publication venues [8–10]. This diversity of data sharing activities and norms is also reflected in the fact that there are different ways in which data is shared and described. For example, Gorgolewski et al. [11] discussed different means of hosting data, including external data repositories, self-hosting, and journal-created repositories. Similarly, Lawrence and colleagues [12] reviewed a few different models of connecting data, its descriptions, and the scientific research that uses the data. From a socio-technical perspective, no data sharing model can support all real-world scenarios. Among all approaches reviewed above, one of the most popular ways to connect scientific publications and datasets is through data citation; this practice has been extensively studied in open science and quantitative science studies literature [13–15].

Beyond data citation, the data availability statement (DAS)—an author-supplied statement in the article's paratext specifying whether the data supporting the research is shared and if so, how it can be accessed—has increasingly been used by researchers to offer information about research data during the past few years. The Public Library of Science (PLOS) was one of the first publishers to implement a DAS policy in the beginning of 2014 [16]; it was followed by such other major publishers as Springer Nature journals [17] and the American Meteorological Society [18]. This approach to data sharing has been extensively embraced because it offers more granular information about research data, which can help researchers better understand the nature of the shared data.

As more publications have been published with a DAS, the statement itself is becoming a novel data source that can shed fresh light on data sharing behavior. However, very few empirical studies have been conducted to examine what information is included in these statements. One notable exception is Federer and colleagues' work [19], which investigated the extent to which the data availability policy was complied with in all PLOS ONE research articles published between March 2014 and May 2016. The authors examined the data sharing mechanisms and repositories used in the research and reported that fewer than 20% of all publications share their data on external data repositories. The value of this paper, however, is overshadowed by the fact that it only covers a very short publication window, from which it is difficult to observe meaningful trends over time.

To fill this gap, this paper presents a larger-scale dynamic study on the contents of data availability statements. Specifically, we examined the different ways in which data is shared and data repositories are used, based on the data availability statements in research articles published in the journal *PLOS ONE* from 2014 to 2020, effectively expanding the research framework used by Federer et al. The following research questions are pursued:

**RQ1: What data sharing mechanisms are mentioned in the data availability statements of *PLOS ONE* articles?**

In this work, we strive to understand how the different data sharing mechanisms are mentioned in data availability statements. By slightly revising the data sharing mechanism ontology offered by Federer et al. [19], we are able to evaluate how these different mechanisms are used over time and across different knowledge domains covered by the journal. In particular, given the arguments that external data sharing (i.e., using a data repository) is more effective than sharing the data in the paper or its supplementary materials [19] and that "upon request" is a profoundly insufficient data sharing mechanism [20,21], our results will aid in understanding whether the general quality of research data sharing has improved during the past few years.

**RQ2: What data repositories are mentioned in the data availability statements of *PLOS ONE* articles?**

Data repositories are an increasingly important component of research infrastructure under the open science movement [22], and information science researchers are paying increasing attention to the roles played by these repositories in scholarly communication [23–26]. However, the above case studies are unable to build deeper insights into how data repositories are used over time and across different knowledge domains. To address this gap, we specifically evaluated how individual data repositories are used in our paper sample along temporal and disciplinary dimensions, as a first step towards a more contextualized understanding of the market of data repositories and their use in specific research contexts.

## 2 Literature review

### 2.1 Empirical studies on data availability statements

The 21st century marks the rise of the data-driven research paradigm [27], which requires that data be effectively shared to facilitate its reuse and enhance the reproducibility and transparency of science [28]. During the past decade, many stakeholders, including funding agencies and publishers, have enforced data sharing policies. The inclusion of a data availability statement is one policy increasingly embraced by academic communities as a means of data sharing [16,17,29].

As a result of these policies, the number of publications with a data availability statement has been increasing over the past few years. Federer et al. [19] extracted all data availability statements from 47,593 articles published in *PLOS ONE* between 2014 and 2016 and reported the expected increase in the number of statements. Graf et al. [30] analyzed statements from 176 Wiley journals between 2013 and 2019 and saw a similar trend across all knowledge domains. Despite the increasing numbers, however, empirical evidence has suggested that this policy per se is insufficient to guarantee the accessibility of research data [31,32].

One factor that prevents the statement from totally fulfilling its goals in scholarly communication is that it may contain different data sharing mechanisms selected by researchers in their specific research contexts. Federer and colleagues (2018) identified 10 major examples of such mechanisms; these ranged from providing data upon request, to sharing it in the paper and/or supplementary materials, to using a data repository. The upon-request approach has been regarded as deeply problematic given that many researchers do not respond to such requests sent by others [20,21]. It has also been argued that sharing data within the paper is insufficient because the data shared in this space tends to be summary data instead of the raw data that researchers used [19]. Data repositories, meanwhile, have been recommended by many publishers because they can provide long-term storage and persistent access to research data [33], partly thanks to the provision of DOIs to datasets in many of these services. However, based on existing works, such repositories are far from the dominant data sharing method [19,34].

Despite these limitations, the increasing use of data availability statements makes them an emerging data source useful for understanding how research data is shared in publications. Nonetheless, most existing studies on this topic aim only to assess the extent to which such statements are available and what data sharing mechanisms are used in a general corpus. No study has yet been conducted to locate data sharing activities within temporal and disciplinary frameworks. This is a major motivation of the present work.

**2.2 How are data repositories used to share data?**

Data repositories are a central instrument in data sharing activities [35,36]. Plantin et al. [37] stressed how data repositories, as exemplified by Figshare, help to reconfigure the logic of our traditional, paper-based scholarly communication system and introduce research data into this system. There is also empirical evidence that data repositories positively affect data search and sharing behavior [38–40] and that research with data deposited in repositories can receive more citations than research without [41].

As more journals implement data sharing policies, there are increasingly many data repositories for researchers to choose from. The re3data (Registry of Research Data Repositories) project[1], one of the most comprehensive registries for research data repositories, indexes over 3,000 repositories in its system. These vary in many aspects, including discipline, scope, file size, and type of data [42–44]. Biomedical sciences are particularly reliant upon data repositories [45], many of which are components of the National Center for Biotechnology Information (NCBI) data infrastructure. More recently, a few interdisciplinary data repositories, such as Open Science Framework (OSF) and Figshare, have been established, as data science grows more mature and spreads to more research fields [42]. Despite the interdisciplinary nature of research data [46], the distinction between thematic and interdisciplinary data repositories has been useful for understanding the roles these repositories play in scholarly communication [44,47].

As data repositories become key mediating agents between publications and research data, it is vital to understand how they are used by researchers, especially from a quantitative perspective. Thus far, only a few studies have tried to analyze individual data repositories from this perspective [23,24,26], and these are not able to provide an overview of the whole field. This issue, however, can be solved by the increasing prevalence of data availability statements, as data repositories are a major type of content recorded in these statements (Federer et al., 2018). This project thus aims to evaluate the use of repositories, particularly over time and across knowledge domains, as a first step towards a deeper appreciation of the roles played by data repositories in our scholarly communication system.

**3 Methods**

---

[1] https://www.re3data.org

## 3.1 Data collection

We retrieved all 228,505 English-language research articles published in *PLOS ONE* up to the end of 2020 in XML format using the *rplos* package of the R programming language [48]. We parsed the data availability statements as well as other metadata elements from the downloaded XML files. *PLOS ONE* was selected as the data source because 1) it is one of the first journals to implement a data availability statement policy (since March 2014); 2) it is one of the mega-journals including all disciplines, allowing us to examine disciplinary variations; and 3) it allows us to compare our results with those from Federer et al. (2018) that rely on the same journal.

Most *PLOS ONE* papers published no earlier than 2014 have a data availability statement (as shown in Table 3 in section 3.4) as a consequence of the journal policy. Given the centrality of this statement to the present study, only the 127,935 articles (55.99% of all articles) with a data availability statement were included in the final sample.

## 3.2 Identification of the data sharing mechanism and repository mentioned in the statement

We extracted data sharing mechanisms and repositories from all selected data availability statements using regular expression matching. To increase the accuracy of our method, all statements were parsed into sentences using the NLTK Python package [49], before algorithms were applied to them. Sentence-level information was later combined at the paper level. The two steps, i.e., extracting the data sharing mechanism and data repositories, are discussed below. The algorithm we used has been shared on Figshare [50].

**We first extracted information regarding the data sharing mechanism.** We simplified the research framework proposed by Federer and colleagues (2018) to include the six categories listed in Table 1. For each category (except *Combination*), all sentences were given a True or False value based on whether certain cue terms or phrases were identified in each sentence. Cases in which two or more categories exist in the same statement (on either the sentence or the statement level) were coded as *Combination*.

We did not examine the other four categories in Federer et al.'s study (including *Access Restricted*, *Location Not Stated*, *N/A*, and *Other*) because (1) they are used much less frequently than the six included categories and (2) some of these categories, especially *N/A* and *Other*, are reflected in those papers that are not assigned any category in our scheme, given the different design adopted in our analysis (i.e., our scheme is not intended to positively categorize every item in the sample).

**Table 1: Classification of data sharing mechanisms in our study**

| Category | Definition from Federer et al. (2018) | Sample Statement from Federer et al. (2018) |
| --- | --- | --- |
| In paper | Statement indicates data are reported in the paper, including in tables and/or figures | All relevant data are within the paper. |
| In SI | Statement indicates data are reported in the Supplemental Information | All data can be found in the Supporting Information. |

| In paper and SI | Statement indicates data are reported in both paper and Supplemental Information | All relevant data are within the paper and its Supporting Information files. |
|---|---|---|
| Repository | Statement names a publicly accessible location where the data are available, such as a repository or website | Data are accessible in <repository>. |
| Upon request | Statement says that author or other individual or group must be contacted to access data[2] | Data made available to all interested researchers upon request. |
| Combination | Statement mentions more than one mechanism for sharing | Sequencing data is available from <repository>. All other data is available in supplementary information. |

**For those papers whose data is shared through one or more repositories, we identified which repositories are mentioned in each paper.** We focused on 89 specific repositories that include: 1) the twenty most frequently mentioned repositories identified in Federer and colleagues' study (2018) and 2) all repositories recommended by *PLOS ONE*[3]. For each category, we searched its name variations to identify name mentions from the statement. We did not consider any *institutional repository* or *non-repository website* treated as an individual repository in the Federer work, because the identity of these repositories or websites is difficult, if not impossible, to extract from the text, even though these cases were found to be heavily used in Federer et al. (2018). Another repository in the Federer list that we did not consider is "National Center for Biotechnology Information," as this often denotes the whole NCBI infrastructure instead of a single data repository. Based on the number of papers with these three "repositories" provided in the Federer paper (Table 2), we expect that our *Repository* category misses about 10% of papers that could have been identified using their methodology.

It should be mentioned that researchers frequently use web links and DOIs to refer to a repository, where the repository name is not necessarily mentioned in the statement. To address this scenario, we also extracted all URLs, including those embedded into DOIs (whether as a link or not) and Handle links (i.e., links starting with "hdl.handle.net"). Cases of the latter two types were common in our sample, which do not show the identity of the web page. Where possible, we resolved them to obtain the repository link. We did not include any DOI or Handle link that could not be resolved.

For all links extracted from the statement texts, DOIs, and Handle links, we extracted the hostname to evaluate whether a link belongs to a specific repository. We created a confusion matrix between repository names and the hostname when both types of information were available. Using the matrix, we identified all hostname-repository pairs appearing at least twice in our data and used such links to classify statements where there was only an identified link but no repository name.

---

[2] In this study, even though we retained the original definition of *Upon Request* in the Federer work, we found that our operationalization of this category at least partially covers Federer et al. (2018)'s *Access Restricted* category, based on a manual inspection of their raw data.

[3] https://journals.plos.org/plosone/s/recommended-repositories

To ensure the accuracy and reliability of the automatic extraction, one coder validated the results by manually inspecting 200 randomly selected data availability statements (evenly split between those with and without any identified repository). The results were used to update our regular expression patterns. This step was repeated until no major mistake could be identified from the results. In our last validated batch, the accuracy for the data sharing mechanism is 96%, and that for data repositories is 97%.

### 3.3 Data analysis

For analysis that focuses on individual repositories, we selected only the 10 repositories most frequently mentioned in our sample so that the sub-sample size would be large enough to show meaningful patterns. We followed the classification system proposed by Pampel and colleagues (2013), which includes *multidisciplinary*, *disciplinary*, and *institutional repositories*. As institutional repositories are not included in the scope of the present study, we primarily focused on the first two categories. For example, Gene Expression Omnibus (GEO) is a repository for microarray data and GenBank for gene sequences, whereas Figshare and Dryad welcome data from multiple domains. This system is applied to the top 10 data repositories we identified from our paper sample, and the result of the classification, as well as the founded year and whether a repository automatically assigns DOIs to datasets, is presented in Table 2.

Table 2: Top 10 Data Repositories in our sample

| Repository | Year Founded | Classification | Providing DOI |
|---|---|---|---|
| Figshare | 2011 [23] | Multidisciplinary | Yes |
| GEO | 2000 [51] | Disciplinary | No |
| Dryad | 2010 [52] | Disciplinary | Yes |
| GenBank | 1982 [53] | Disciplinary | No |
| SRA | 2009 [54] | Disciplinary | No |
| OSF | 2012 [55] | Multidisciplinary | Yes |
| GitHub | 2008 [56] | Multidisciplinary | No |
| Dataverse | 2007 [57] | Multidisciplinary | Yes |
| Zenodo | 2013 [24] | Multidisciplinary | Yes |
| Bioproject | 2011 [58] | Disciplinary | No |

To understand the disciplinary patterns of data sharing and repository use, we used the paper-level subject field classification developed by the Centre for Science and Technology Studies (CWTS) at Leiden University (*knowledge domain* hereafter), one that has been proven to be an accurate classification instrument in previous studies [59,60] and has been applied in the practice of the *Leiden Ranking*[4].

Using the top-level categories in this scheme, all *PLOS ONE* publications were classified into one of following five knowledge domains: Biomedical and Health Sciences (BHS), Life and Earth Sciences (LES), Social Sciences and Humanities (SSH), Mathematics and Computer Science (MCS), and Physical Sciences and Engineering (PSE). It should be noted that not every paper was assigned a domain in this scheme. Therefore, for domain analysis, we considered only the 125,425 articles that had any category in CWTS' in-house database.

**3.4 Descriptive analysis**

Table 3 summarizes the absolute and relative numbers of papers with a data availability statement per year. Although PLOS' data availability policy was implemented in 2014, only in the following year did most publications in *PLOS ONE* have such a statement.

**Table 3: Yearly distribution of all publications and those with DAS**

| Year | Papers with DAS | Papers in the journal | Percentage of papers with DAS |
|---|---|---|---|
| 2014 | 9,596 | 27,161 | 35.33% |
| 2015 | 27,318 | 28,036 | 97.44% |
| 2016 | 21,463 | 21,507 | 99.80% |
| 2017 | 20,387 | 20,389 | 99.99% |
| 2018 | 17,878 | 17,878 | 100.00% |
| 2019 | 15,290 | 15,290 | 100.00% |
| 2020 | 16,003 | 16,003 | 100.00% |

Table 4 describes the number of papers with data availability statements in each knowledge domain. In our sample, more than 87% of papers are from biomedical and health sciences or life and earth sciences.

**Table 4: Distribution of all publications with DAS across knowledge domains**

| Knowledge domain | Number of papers with DAS | Percentage (*n* = 125,425) |
|---|---|---|
| Biomedical and Health Sciences (BHS) | 83,906 | 66.90% |
| Life and Earth Sciences (LES) | 25,281 | 20.16% |
| Social Sciences and Humanities (SSH) | 7,850 | 6.26% |
| Mathematics and Computer Science (MCS) | 4,259 | 3.40% |

---

[4] See more information about the subject field classification system applied in the Leiden Ranking at: https://www.leidenranking.com/information/fields.

| Physical Sciences and Engineering (PSE) | 4,129 | 3.30% |

To further understand how disciplinarity contributes to the changing patterns in the use of data sharing mechanisms and repositories, we plotted the contributions of the above disciplines to our final sample over time. Figure 1 shows an increase in publications from social and computer sciences and humanities in *PLOS ONE*, as compared to hard sciences, since 2014. It shows, as well, that the disciplinary distribution of publications in this journal has been becoming more even over time.

**Figure 1: Percentage of publications in each knowledge domain over time**

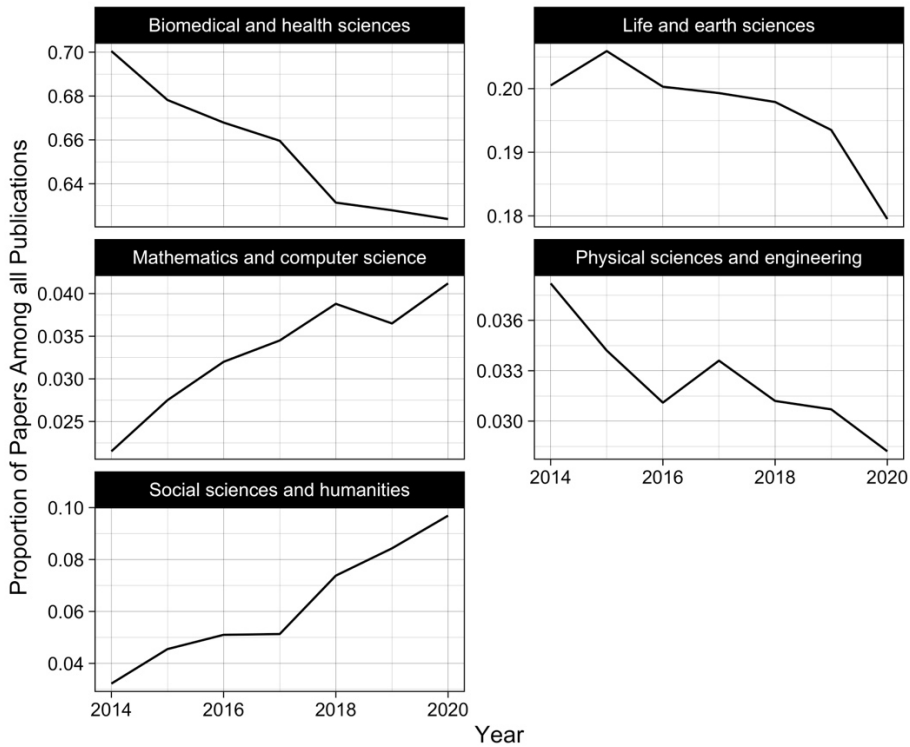

## 4 Results

### 4.1 Data sharing mechanisms as recorded in data availability statements

Building upon our data sharing mechanism classification scheme, we analyzed how the six major mechanisms are used in our paper sample. Table 5 shows the relative frequency of each mechanism in our study as well as in Federer et al. (2018) for comparison.

**Table 5: Percentage distribution of data sharing methods in our study and Federer's work**

| Data sharing mechanism | Our study ($n = 127{,}935$) | Federer's study ($n = 47{,}593$) |
|---|---|---|
| In paper and SI | 48.20% | 45.34% |
| In paper | 17.15% | 24.3% |
| Repository | 13.62% | 15.4% |
| Combination | 4.68% | 4.5% |
| Upon request | 5.05% | 1.4% |
| In SI | 2.14% | 1.4% |

Most of the six categories are used relatively similarly between our work and the previous study. One category that poses a particularly stark difference is *Upon Request*. Through a manual examination of our results, we found that one reason for this large difference is that our category covers many examples classified as *Access Restricted* in the Federer paper. Additionally, about 67.50% of all statements in our study state that the data are shared only within the paper and/or in the supplementary materials, which indicates that these are still the dominant methods adopted in *PLOS ONE*, followed by the *Repository* category. It is not surprising that *Repository* in our results has a lower ratio of publications than in the Federer work, given the much more limited scope of our inclusion criteria. As a result, this number by itself may not fully reflect how repositories were actually used in the journal.

To better understand how data sharing activities change over time, we plotted the ratios of publications with the above mechanisms (except for the *Combination* category) over the years. As illustrated in Figure 2, we found an increasing trend in the use of repositories only to share research data in *PLOS ONE* articles, from about 10.35% in 2014 to 16.63% in 2020. By comparison, paper and supplementary materials mechanisms are generally used much less frequently over time. Despite the individual differences, when we consider all statements using internal sharing ("Paper & SI Total," i.e., those sharing data in the paper or supplementary materials), the percentage of publications decreased from 73.44% to 60.28% during the publication window that we examined. This shows the contrasting patterns in internal versus external data sharing, even though the rise of the latter does not totally compensate for the decrease in the former.

**Figure 2: Usage trends in different data sharing mechanisms over time**

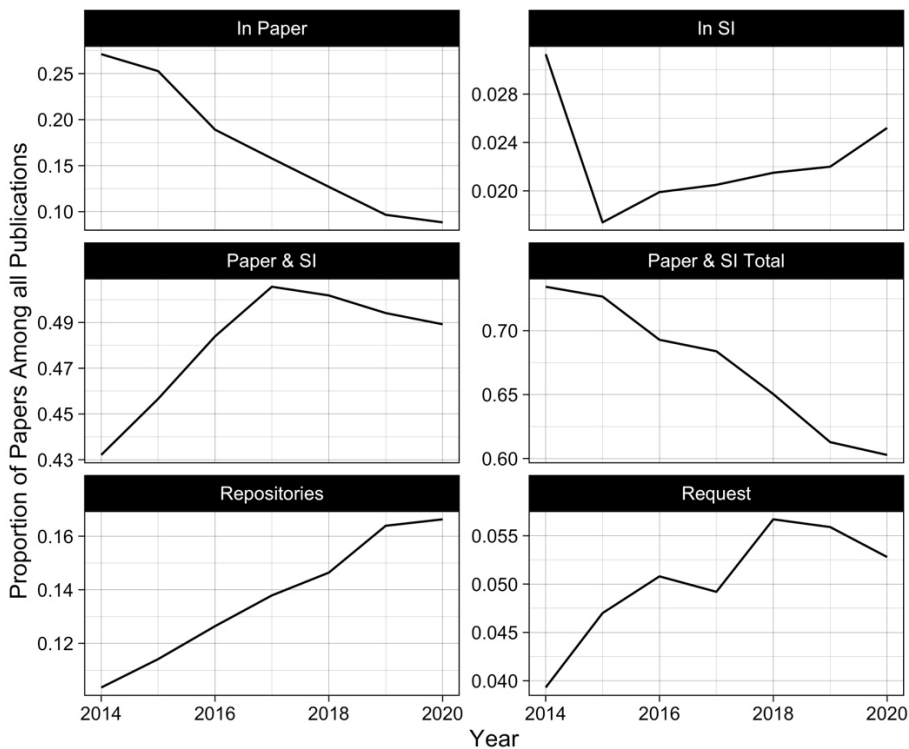

Figure 3, in turn, summarizes how the six data sharing mechanisms are used across the five knowledge domains in the CWTS classification scheme. The graph shows that there are very different data sharing patterns across domains. Physical sciences and engineering (PSE) and biomedical and health sciences (BHS) are relatively similar to each other in that they are more heavily reliant upon sharing data in the paper and/or supplementary materials, whereas social sciences and humanities (SSH) as well as mathematical and computer sciences (MCS) are more strongly connected to data repositories. Such differences indicate that there may be distinct community norms regarding data sharing across knowledge domains, which may be attributed to the research culture as well as the nature of the research and data.

**Figure 3: Data sharing mechanism usage across knowledge domains**

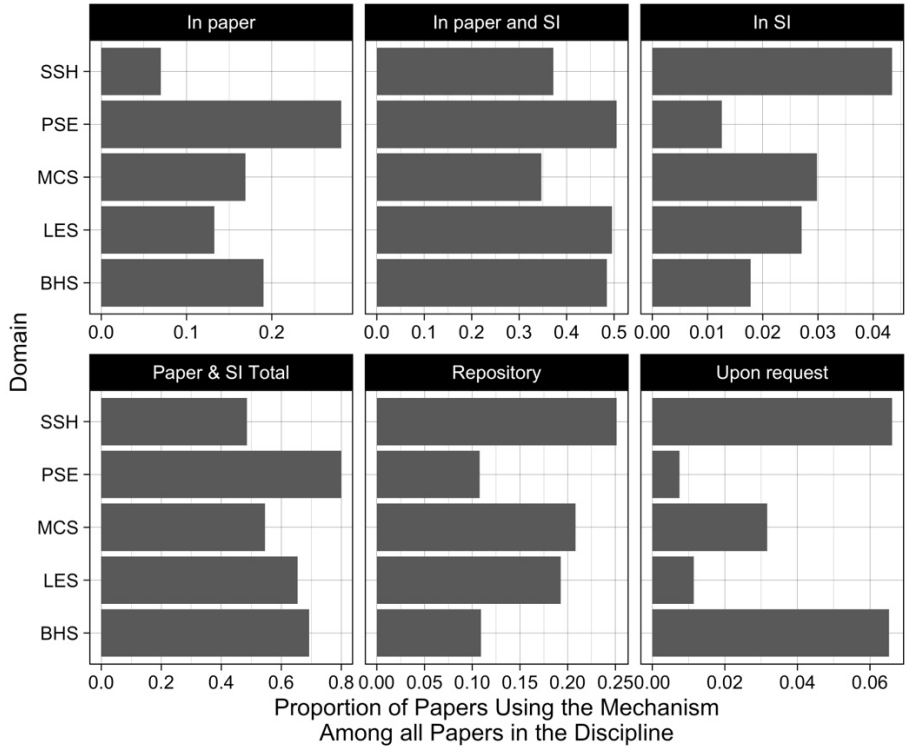

To understand the extent to which disciplinarity contributes to the shift in data sharing mechanisms, we plotted the information in Figure 3 over the timeline, as shown in Figure 4. The y-axis of the graph shows the proportion of articles in each knowledge domain that use a specific data sharing mechanism over all publications in the same domain (for example, in the domain of PSE, there have been 70% to 80% of articles that share data in the paper or SI over time). The graph shows that each mechanism exhibits relatively parallel changes across the knowledge domains, showing that the major trends described in Figure 2 are playing out in every knowledge domain covered by the journal.

**Figure 4: Temporal shift in data sharing mechanism usage by discipline**

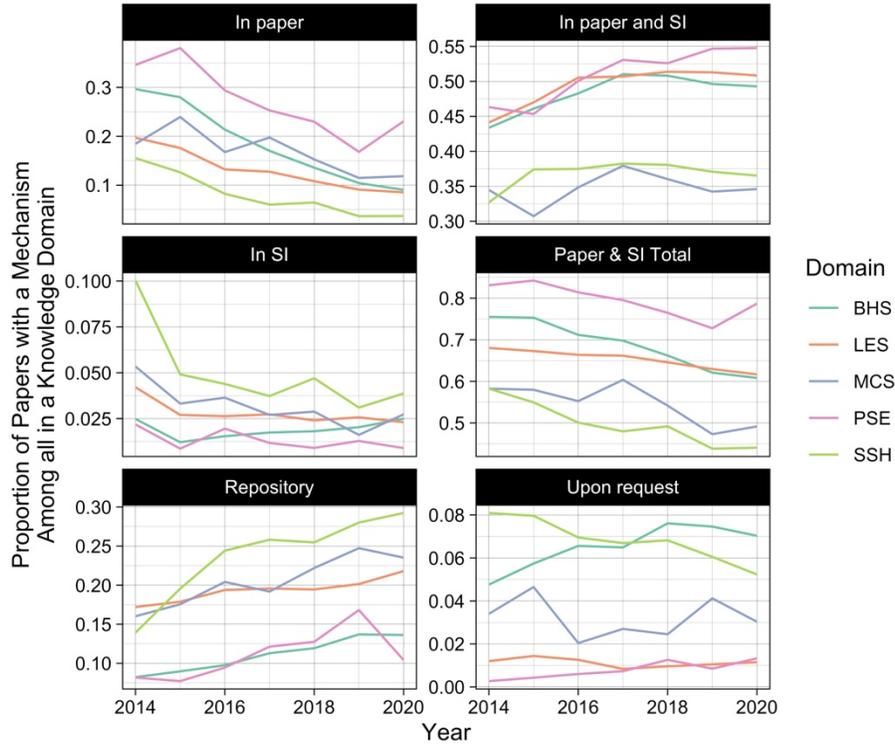

## 4.2 How are individual repositories mentioned in DAS?

Of the 89 data repositories examined in this research, we found 72 being mentioned in our sample: they are mentioned 25,831 times in 22,356 statements. Among these statements, 3,057 (13.67%) mentioned more than one data repository. Table 6 summarizes the 10 most frequently mentioned data repositories in our paper sample; these are mentioned in more than 80% of all statements that use a repository as a data sharing method, whereas nearly 80% of data repositories are each mentioned in fewer than 1% of publications.

Table 6: How are top 10 data repositories used?

| Repository | Count | Percentage of all Repository Mentions ($n = 25,831$) | Percentage in Federer et al. ($n = 8,702$) |
|---|---|---|---|
| Figshare | 4,771 | 18.47% | 16.62% |
| GEO | 2,955 | 11.44% | 11.50% |
| Dryad | 2,759 | 10.68% | 11.34% |
| GenBank | 2,687 | 10.40% | 11.48% |
| SRA | 2,057 | 7.96% | 7.37% |
| OSF | 1,937 | 7.50% | 1.40% |
| GitHub | 1,906 | 7.38% | 3.22% |
| Dataverse | 1,026 | 3.97% | 2.49% |
| Zenodo | 944 | 3.65% | 1.15% |
| Bioproject | 914 | 3.54% | 0.91% |

We examined how these 10 repositories are used over time. A few patterns emerge from Figure 5. First, there is a broad pattern that more general (multidisciplinary) repositories, most of which have relatively short histories and can assign DOIs to datasets, are increasingly used during the past decade. In contrast, those repositories dedicated to life and biological sciences tend to be decreasingly used, with the sole exception of Bioproject, which may be attributed to the fact that (1) Bioproject is the most recent member in this category based on the information in Table 2 and (2) Bioproject, as a special data repository, is frequently co-used with other NCBI repositories in the list. Second, due to these individual changes, the top 10 repositories compose a stable percentage of all repository mentions in our sample. The number has stayed between 83–84% during the seven-year publication window.

**Figure 5: How are each of the top 10 repositories used over time?**

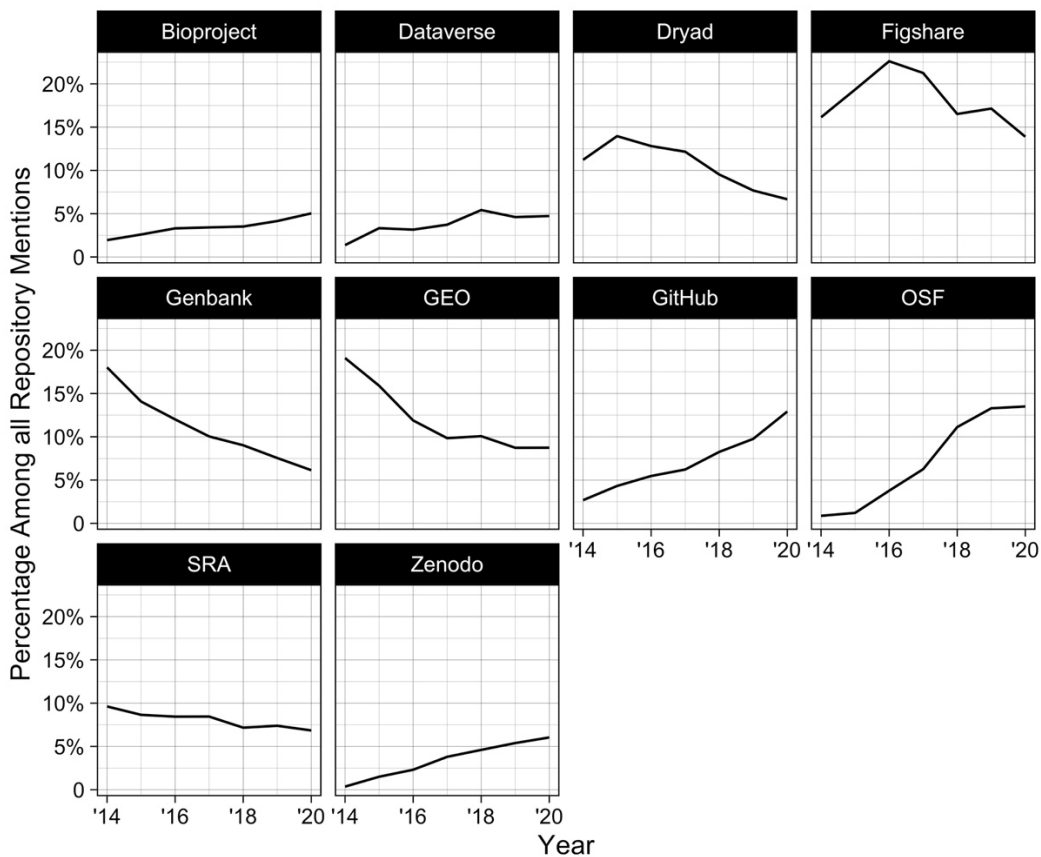

### 4.3 How are repositories used different across knowledge domains?

We next examined the percentage of DAS with at least one repository mentioned in each knowledge domain. We found that social sciences and humanities, life and earth sciences, and mathematics and computer science are much more likely to use DAS than biomedical and physical sciences. This indicates that there are distinctly different data sharing cultures across different knowledge domains, which may be explained by their unique relationships with data and the data-driven research paradigm.

**Table 6: Percentage of DAS mentioning any data repository across knowledge domains**

| Domain | Repository mentions | Percentage within domain |
|---|---|---|
| Social sciences and humanities | 2,225 | 28.34% |
| Life and earth sciences | 6,470 | 25.59% |
| Mathematics and computer science | 1,077 | 25.29% |
| Physical sciences and engineering | 600 | 14.53% |
| Biomedical and health sciences | 11,869 | 14.15% |

Figure 6 further summarizes how the top 10 repositories are used in the disciplinary context, whose x-axis represents the proportion of articles that uses a specific data repository over all publications in a knowledge domain. By tracing the number of papers mentioning each repository in a specific domain, the graph shows that the usage pattern of these repositories can be largely separated along the line between multidisciplinary and disciplinary repositories.

**Figure 6: How are the top 10 repositories used across knowledge domains? (Multidisciplinary repositories shown in blue and disciplinary repositories in red)**

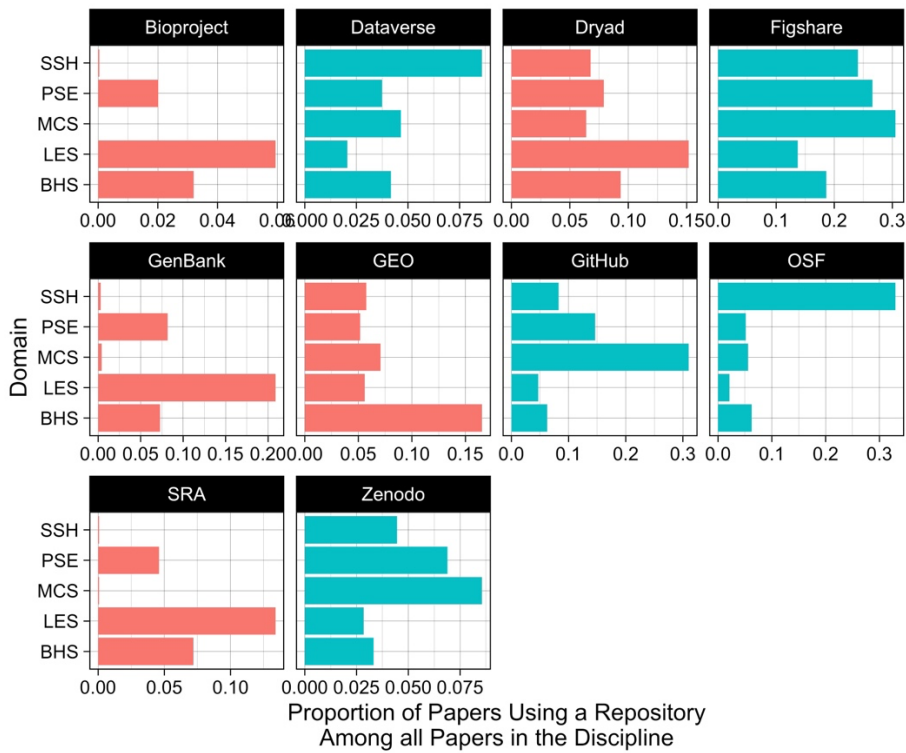

Most of the disciplinary repositories, such as Bioproject, Dryad, GenBank, GEO, and SRA, are predominantly used by either LES or BHS, two fields that are tightly focused on the research interest of the journal. It is also clear that Dryad and GEO are more likely to be used in SSH and MCS than the other three repositories discussed above, indicating that they may contain data that are more general in scope.

The five multidisciplinary repositories, by comparison, show very different usage patterns: all of them are used more heavily in the first three domains (i.e., SSH, PSE, and MCS) than in LES and BHS, showing their deeper connections to broader disciplinary contexts. In particular, both OSF and Dataverse are primarily used in SSH. The relationship between Dataverse and social science is well documented in the literature [61], but the very strong usage of OSF in this domain will need future research to offer a better explanation. Moreover, as a service that is deeply connected to programming, it is not surprising that GitHub is most frequently used in MCS, even though in this work, we did not distinguish another important function played by GitHub, i.e., sharing code, from general data sharing practices.

We plotted the changing usage pattern of each repository in the five knowledge domains to further contextualize the patterns identified above. Figure 7 shows the results, with the y-axis representing the percentage of publications mentioning the repository out of all repository mentions in the domain. In this figure, repository-domain pairs that exist for fewer than six years in our sample are removed (most of these pairs have very few publications, such as SRA-SSH in 2016).

**Figure 7: Usage trends of individual data repositories across knowledge domains**

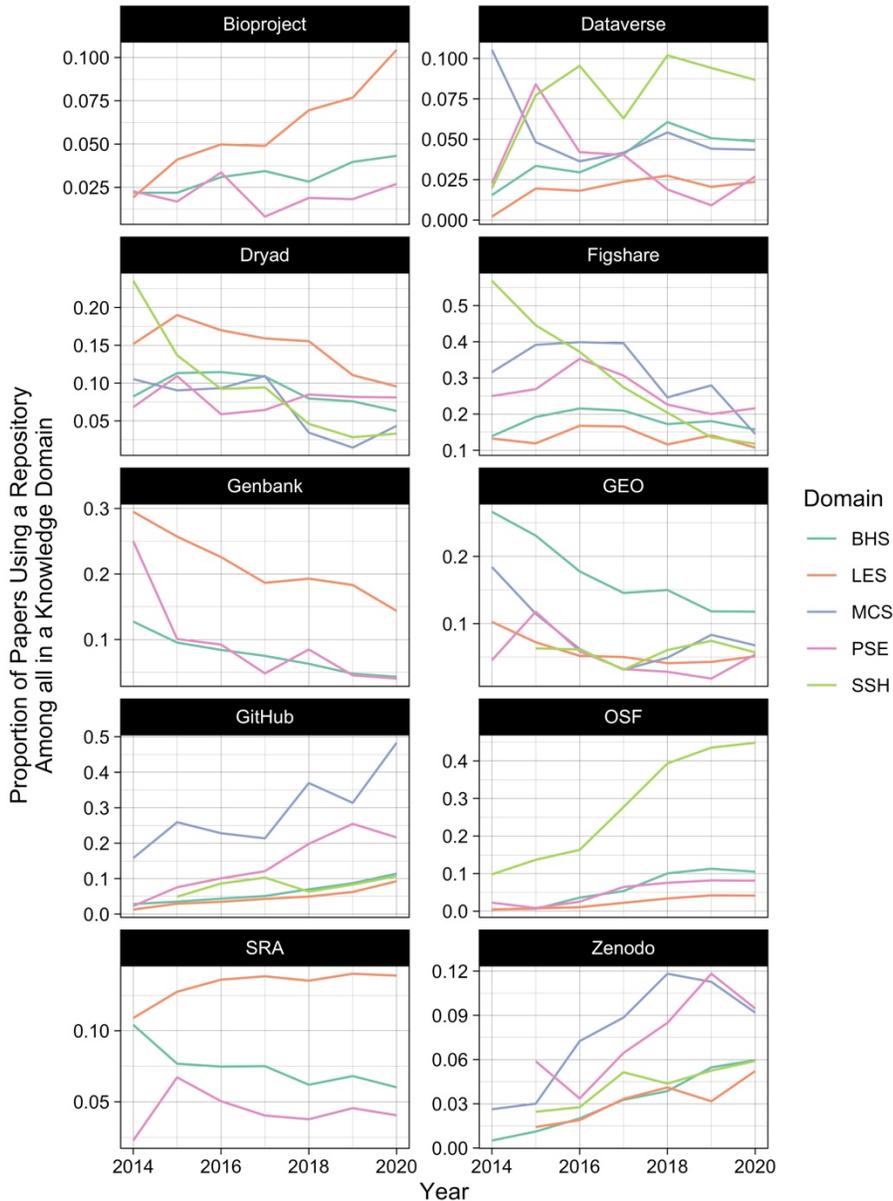

The graph shows that most of the significant changes in the use of a repository follow similar patterns, from the most dominant domain to relatively minor ones. For example, OSF, GitHub, GenBank, and GEO each have a dominant domain, and their changes over time can be observed in both the dominant and minor domains. It is also interesting that some repositories' dominant domain has changed over time, notably in the cases of Dryad, Dataverse, and Figshare, which sheds light on the fickle nature of the bond between a domain and a data repository (especially a multi-disciplinary repository).

Figure 8 shows the same information in Figure 7 but is organized by domain instead of by repository. In this graph, domain-repository pairs whose yearly mean ratio is lower than 0.1 (i.e., repositories that are used in fewer than 10% of all publications in a domain in a specific year) are removed from the graph to reduce noise. From this domain-centric view, it is obvious that MCS and SSH have particularly strong replacement effects among the top repositories. For example, in the case of MCS, GitHub has clearly replaced Figshare as the dominant repository during the publication window examined in this work, and the paths of the two repositories are essentially mirror images. The same can be said of OSF and Figshare in the domain of social sciences and humanities. BHS, LES, and PSE have more repositories being used, and the usage of data repositories is less centralized in these three domains.

**Figure 8: Usage patterns of top repositories within each knowledge domain**

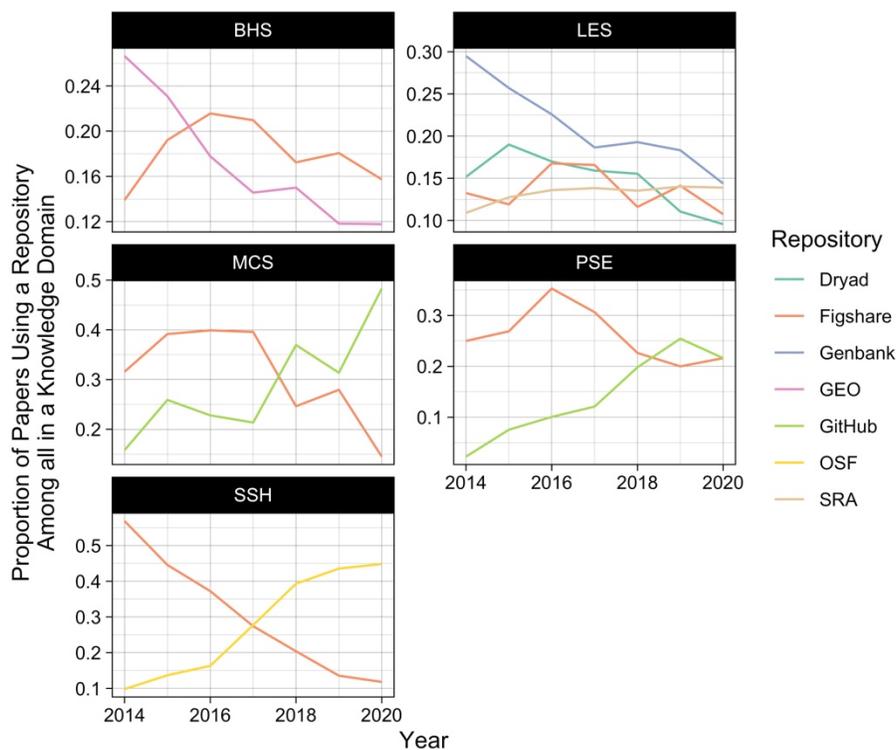

## 5 Discussion

### 5.1 Data sharing mechanisms in data availability statements

In this work, we analyzed the contents of all data availability statements in *PLOS ONE*, with a strong focus on the data sharing mechanism and data repositories being described in the statement. After extracting such information using the regular expression method, we further evaluated how different data sharing mechanisms and repositories are used across knowledge domains and over time, as a first step towards a more contextualized understanding of how research data is shared using the data availability statement, a novel data source that focuses on data sharing.

The first part of our analysis concerns the data sharing mechanism being recorded in the statements. Our findings show that there is not a significant change in how the major mechanisms are used in the journal when we consider the whole sample, as compared to Federer et al. (2018), even though some key categories are going through gradual changes over time. We found that sharing data within the paper and/or the supporting information section is still the dominant mechanism used in our sample. Considered as a whole, these two approaches are used in more than 67% of all publications, though this number declines from more than 75% in 2014 to just above 60% in 2020. As discussed by Federer et al. (2018), a major limitation of this approach to sharing data is that it normally only includes summary data, which is not sufficient to reproduce the original results.

The above factor is an important reason for PLOS' policy of using external data repositories as the top recommended data sharing mechanism. Our results demonstrate that there is a slow yet stable growth in external data repository use among all *PLOS ONE* articles (from about 10% in 2014 to 16% in 2020), which compensates for part of the decrease in internal data sharing. We also found that the above patterns in the use of data/supporting materials and external data repositories over time are largely consistent across the five knowledge domains, despite their individual practices. More specifically, social science and humanities displays particularly strong growth in data repository use, which may be connected to the fact that a few multidisciplinary data repositories, especially OSF and Zenodo, were created in the early 2010s. Our results largely demonstrate that data sharing is becoming better implemented in *PLOS ONE* over time, though more studies will be needed to better understand the bigger picture as well as how data sharing is rooted in specific research communities—particularly those not covered comprehensively by this journal.

Moreover, we found that despite the heavy criticism it has received [20,62,63], sharing data by personal request has been increasingly used in the journal during the publication window examined in this study. Following the suggestions offered by the studies mentioned above, we also advocate for the development of more open data sharing policies by journals and research funders; at the same time, a better infrastructure can be built to support researchers in more effective data sharing.

**5.2 Data repository usage over time and across disciplines**

As external data repositories continue to play increasing roles in data sharing activities, the second part of this work focuses on which repositories are used, along with how they are used over time and across different knowledge domains. We identified the 10 most frequently used data repositories from our paper sample, which are very similar to those reported by Federer and colleagues (2018). Using our data, which come from a larger publication window, we observed interesting temporal patterns in how these repositories are used. In particular, we found that multidisciplinary data repositories, especially GitHub and OSF, tend to be increasingly used over time, whereas most of the disciplinary repositories show a decreasing trend. We strive to examine how the individual histories of the repositories and whether a repository support assigning DOIs to datasets could contribute to this pattern. There is a general pattern that recent repositories are likely to be used more often during our window than their older counterparts, though this finding is inevitably subject to survivor bias and thus should be taken with greater caution. Moreover, there is a strong overlap between multidisciplinary data repositories and those that support DOI assignment. As a result, we cannot safely draw any conclusion about whether DOI-related functions help with the popularity of data repositories.

The finding that disciplinary repositories tend to be used less frequently is a function of the disciplinary composition of publications in our paper sample. Most of the disciplinary repositories covered by this research are focused on the domains of biomedical and life sciences. Consequently, as more research from other domains is published in the journal over time, multidisciplinary repositories enjoy a major advantage. Given this reason, our results cannot be simply used to support the conclusion that certain types of data repositories are gaining momentum over others in the scholarly communication system.

Instead, our results indicate that many repositories have parallel temporal tends across multiple knowledge domains. For example, the decreasing usage of GenBank seems to be evident across various knowledge domains, so does the increasing trend of OSF and GitHub. Interestingly, there is another pattern among the top 10 repositories: some repositories exhibit radical changes in their dominant domains but not in minor ones, as in the cases where Figshare and Dryad have seen a large decrease in use in the social sciences and humanities. Such examples, especially Figshare, can be explained by the replacement effect on the repository level, where its popularity is clearly overtaken by OSF in the domain of social science and humanities (as well as by GitHub in mathematics and computer science). This leads to an interesting question of how the data repository market is shifting over time, particularly under the framework of *entitymetrics* [64], which warrants future study.

## 6 Conclusion

In this work, we used automatic text matching techniques to extract the data sharing mechanisms and data repositories mentioned in 127,935 data availability statements in *PLOS ONE* publications from 2014–2020. We expanded the frameworks developed by Federer et al. (2018) to understand how data is shared in scientific publications, and in particular what and how data repositories are used to share research data. We found increasing usage of data repositories among the examined paper sample, which is consistent with the journal policy. However, sharing data in the paper and/or supplementary materials remains the dominant method, despite its known limitations for scientific reproducibility. In addition, our results show that multidisciplinary data repositories have been increasingly used in the paper sample, primarily driven by the growing degree of multidisciplinarity of the journal. Underlying this general shift, the disciplinary attribute of some (but not all) individual repositories has also undergone significant changes, which implies deeper connections between research data sharing activities and disciplinarity.

This paper offers a novel analysis of data sharing activities inscribed in data availability statements from temporal and disciplinary perspectives. Data availability statements have been receiving increasing attention from researchers during the past few years, as a useful instrument to understand how researchers share data in a more granular manner [17,19,31]. However, very few studies have examined how the information recorded in such statements varies by domain and over time. In this work, by locating data sharing activities in the temporal and disciplinary frames, we demonstrate how data sharing activities and the infrastructure to support these works are approached by researchers dynamically in the setting of *PLOS ONE*. By doing so, the present work bridges an important gap between quantitative science studies and a critical understanding of data-driven research infrastructure [65]. Our results will help to build a dynamic, empirical understanding of how data sharing activities are situated in research practices. This work will also support research journals and funders in rethinking their policies about open science and could furnish important baseline data for future works.

This study, however, suffers from a few limitations that should be addressed in future research. First, this work only focuses on 89 multidisciplinary and disciplinary data repositories, without considering the institutional repositories that also play critical roles in the landscape of data sharing. Second, to interpret repository usage patterns, we only considered temporal and disciplinary factors, yet most of the characteristics of individual data repositories as well as other social, economic, and political factors are at least equally important, such as policies and functions of individual data repositories (especially a repository's easiness of use and the assignment of DOIs to datasets), the nature of data involved in the research, and the affiliation and network of paper authors. In the next step of our project, we will more systematically address such limitations to examine how the selection of data repositories is connected to deeper socio-technical factors in the research system, thereby aiming to offer a more thorough understanding of how data is used and managed to produce scientific knowledge.

research infrastructures. *J Assoc Inf Sci Technol* 2017; 68: 1341–1359.